\documentstyle[epsf,twoside,fleqn,nuclphys]{article}

\newcommand{\be}{\begin{equation}}
\newcommand{\ee}{\end{equation}}
\newcommand{\ba}{\begin{array}}
\newcommand{\ea}{\end{array}}
\newcommand{\bea}{\begin{eqnarray}}
\newcommand{\eea}{\end{eqnarray}}
\newcommand{\seff}{$\sin^2 \theta_{\rm eff}$ }

\newcommand{\th}{\theta}
\newcommand{\kaid}{$\chi^2$/(d.o.f.)\ }
\newcommand{\kai}{$\chi^2$\ }
\newcommand{\zw}{\\ [1 pt]}
\newcommand{\vier}{\\ [4 pt]}

\newcommand{\AmS}{{\protect\the\textfont2
  A\kern-.1667em\lower.5ex\hbox{M}\kern-.125emS}}

\newcommand{\postscript}[2]
 {\setlength{\epsfxsize}{#2\hsize}
  \centerline{\epsfbox{#1}}}

\title{
The SLD Asymmetry in View of the LEP Results
\hfill {\normalsize UPR--0619T} \\ \hfill {\normalsize June 1994}}
\author{Jens Erler\address{University of Pennsylvania,
                         David Rittenhouse Laboratory, \\
                         209 S. 33rd Street,
                         Philadelphia, PA 19104, U.S.A.}}
\begin{document}

\begin{abstract}
LEP determines $\sin^2 \theta_{\rm eff}$, by combining the various asymmetry
measurements to be $0.2321 \pm 0.0005$. On the other hand, the left-right
asymmetry as measured at SLC corresponds to \seff = $0.2292 \pm 0.0010$.
I will discuss the possibilities of this being i) a statistical fluctuation,
ii) a hint for New Physics or iii) an experimental discrepancy.
\end{abstract}

\maketitle

\section{Summary of the experimental situation}

The cleanest determination of the weak mixing angle can be obtained
by measuring various asymmetries. Most systematic uncertainties
cancel out on the Z pole and all asymmetries are very sensitive to \seff .
Moreover, in the context of the standard model, \seff is numerically very
close to $\sin^2 \th_{\overline {\rm MS}}$. A detailed
discussion of LEP asymmetries can be found in~\cite{yellowbook}.
The left-right asymmetry is described in~\cite{SLD1}.

Since near the Z-pole all processes are dominated by the Z-exchange
contributions, it is possible to approximate accurately the relevant
observables by the improved Born formulas. This involves a flavor dependent
definition of the weak mixing angle given by
\be
   \sin^2 \th_{\rm eff}^f = {1\over 4 |Q_f|} (1 - {g_v^f\over g_a^f}),
\ee
where $f$ is the fermion index.
The relevant combination entering the asymmetries is given by
\be
\label{af}
   A_f = \frac{1 - 4 |Q_f| \sin^2 \th_{\rm eff}^f}
   {1 - 4 |Q_f| \sin^2 \th_{\rm eff}^f+ 8 |Q_f|^2 \sin^2\th_{\rm eff}^f}.
\ee
Since $\sin^2 \th_{\rm eff} \approx 1/4$ an accidental cancellation
takes place in the case of $A_e$. On the other hand, all asymmetries are
proportional to $A_e$, so that they are all small and primarily
sensitive to $\sin^2 \th_{\rm eff}^{\rm lept}$.

The forward-backward asymmetries on top of the $Z$-peak, neglecting
$\gamma$ exchange, are given by the combinations
\be
   A_{FB}^0 (f) = {3 \over 4} A_e A_f.
\ee
The resonance Born approximation for the $\tau$ polarization is
\be
    {\cal P}_\tau (\cos\th)=-\frac{A_\tau + A_e \frac{2\cos\th}{1+\cos^2\th}}
                                {1 + A_\tau A_e \frac{2\cos\th}{1+\cos^2\th}},
\ee
so that the average polarization is just given by
\be
   {\cal P}_\tau^0 = - A_\tau.
\ee
${\cal P}_\tau (\cos\th)$ provides~\cite{LEPEWG} nearly independent
determinations of $A_\tau$ and $A_e$,
\be
   {{\cal P}_\tau^{FB}}^0 = - {3 \over 4} A_e.
\ee

More recently, ALEPH, DELPHI and OPAL also measure the average forward-backward
charge asymmetry, $<Q_{FB}>$. It is assumed that fermion dependent vertex
corrections are small. They are assigned to their SM values, so that
$<Q_{FB}>$ can be converted to values of $\sin^2\th_{\rm eff}^{\rm lept}$.
Only {\em after\/} this is done, are the various measurements combined.

At present, only SLC determines the initial state left-right asymmetry
whose Born expression is
\be
   A_{LR}^0 = A_e.
\ee
The SLD group also provides a measurement of the forward-backward polarization
into $b$-quarks,
\be
   {A_{FB}^{\rm pol}}^0 (b) =  {3 \over 4} A_b.
\ee

The results for LEP shown below correspond to the ones quoted
by A.\ Blondel~\cite{Blondel}. As for the left-right asymmetry
see the contribution by R. King~\cite{King} and reference~\cite{SLD2}. The
value of $A_{FB}^{\rm pol} (b)$ is as quoted by B. A. Schumm~\cite{Schumm}.
\be
\ba{rclcl}
A_{FB}(e)  &=& 0.0158 &\pm& 0.0035, \\
A_{FB}(\mu) &=& 0.0144 &\pm& 0.0021, \\
A_{FB}(\tau)&=&0.0221 &\pm& 0.0027, \\
A_{FB}({\rm b})  &=&0.097  &\pm& 0.0045, \\
A_{FB}({\rm c})  &=&0.072  &\pm& 0.0106, \\
A_\tau ({\cal P}_\tau)  &=&0.150  &\pm& 0.010, \\
A_e   ({\cal P}_\tau)   &=&0.120  &\pm& 0.012, \\
\sin^2\th_{\rm eff}^{\rm lept} (Q_{FB}) &=&0.2320 &\pm& 0.0016, \\
A_{LR}     &=&0.1656 &\pm& 0.0076, \\
A_b(A_{FB}^{\rm pol} (b)) &=& 0.99 &\pm& 0.13.
\ea
\ee
The leptonic FB asymmetries have a small correlation which will not
be taken into account in the following. There may also be a correlation
between $A_{FB}(q)$ and the charge asymmetry which is ignored, as well.

A common fit to the LEP asymmetries gives
\be
  \sin^2\th _{\rm eff}^{\rm lept} (A^{\rm LEP}) = .2321 \pm .0005 ,
\ee
with a \kaid of 11.2/7. This value of
$\sin^2\th _{\rm eff}^{\rm lept}$ corresponds to a top mass of
$176 \pm 15$ GeV. Throughout I will assume a Higgs mass of 300 GeV
and $\alpha_s (M_Z) = 0.118$.

The left-right asymmetry as measured at SLC corresponds to
\be
  \sin^2\th _{\rm eff}^{\rm lept} (A^{\rm SLC}) = 0.2292 \pm 0.0010 ,
\ee
and $m_t = 255_{-24}^{+22}$ GeV. Inclusion of the 1992 data changes the central
values to 0.2294 and 249 GeV, respectively.

If the validity of the Standard Model is assumed, then
the measurements of the $Z$ line shape and partial decay rates allow
for another determination of the top mass,
\be
  m_t = 156^{+21}_{-23}\; {\rm GeV},
\ee
corresponding to
\be
   \sin^2\th _{\rm eff}^{\rm lept} (\Gamma^{\rm LEP}) = 0.2327 \pm 0.0007 .
\ee
The combined LEP result is
\be
\label{LEPtop}
\ba{rcrcl}
   \sin^2\th _{\rm eff}^{\rm lept} ({\rm LEP}) &=& 0.2323 &\pm& 0.0004 , \vier
                                m_t^{\rm LEP}  &=& 169 &\pm& 12\; {\rm GeV},
\ea
\ee
which is in excellent agreement with the value of $m_t = 165^{+13}_{-14}\; GeV$
as quoted by A. Blondel~\cite{Blondel}. The slightly smaller error
in~(\ref{LEPtop})
is due to neglecting small correlations. Most of the difference in the central
values arises from a shift of about 3 GeV in the top mass, when top threshold
effects are included in $\Delta \rho$. These effects are
included in the analysis leading to the values in~(\ref{LEPtop}), but they
are not yet fully accounted for in the package ZFITTER which is used by
the LEP groups. Our calculation of $\sin^2\th_{\overline {MS}}$ from $m_Z,
m_t, m_H$ and $\alpha_s$ uses a routine written by B. Kniehl based
on reference~\cite{FKS}. For the relation between $\sin^2\th
_{\rm eff}^{\rm lept}$ and $\sin^2\th_{\overline {MS}}$ we use the analysis
of P. Gambino and A. Sirlin~\cite{Sirlin}.

Combining the most recent results from CDF, D0 and UA2 yields a $W$ mass of
\be
  m_W = 80.23 \pm 15\; GeV
\ee
which corresponds to
\be
\ba{rcl}
     \sin^2\th _{\rm eff}^{\rm lept} (m_W)& =& 0.2326 \pm 0.0008 , \vier
                                 m_t^{m_W}& =& 162^{+21}_{-26}\;\;  {\rm GeV}.
\ea
\ee
Low energy neutrino nucleon scattering data give a somewhat lower central
value for the top mass, viz.
\be
\ba{rcl}
     \sin^2\th _{\rm eff}^{\rm lept} (\nu N)& =& 0.2333 \pm 0.0015 , \vier
                              m_t^{\nu N}& =& 137^{+48}_{-70}\;\;  {\rm GeV}.
\ea
\ee
Finally, the candidate top events at CDF~\cite{CDF} suggest
\be
\ba{rcrcl}
                 m_t^{\rm CDF} &=& 174 &\pm& 16\;\;  {\rm GeV}, \vier
     \sin^2\th _{\rm eff}^{\rm lept} ({\rm CDF}) &=& 0.2322 &\pm& 0.0005 .
\ea
\ee

Figure~\ref{xxmt5} illustrates the described situation and shows
$\sin^2\th_{\overline {MS}}$ as a function of $m_t$.
$M_Z = 91.1895 \pm 0.0044$ GeV is treated as an independent measurement
rather than an input. The lower top mass limit from D0~\cite{D0} of 131 GeV
(95\% c.l.) is also shown. The 90\% confidence region is with respect to
the indirect data excluding SLC.

\begin{figure*}[htb]
\vspace{9pt}
%
\makebox[160mm]{\postscript{xxmt5.ps}{0.87}}

\caption{1 $\sigma$ contours of NC observables in the
$\sin^2\th_{\overline {MS}}$ -- $m_t$ plane and combined
90\% confidence region.}
\label{xxmt5}
\end{figure*}

Table~\ref{tabfits} shows various combined fits to the data sets described
above. The \kai value of a fit can be converted into the ``goodness of the
fit'' which is defined as the probability that a randomly chosen set of data
gives a higher value of \kai .

When leaving out the left-right asymmetry the goodness of the fits
is generally satisfactory. It is worst for the LEP asymmetries which are
consistent with the Standard Model only at the 13.1\% c.l. and it is best when
all indirect data as well as the CDF result are included in which case
the fit shows
consistency at the 34.3\% c.l. It should be kept in mind, however, that
these remarks depend slightly on the choice for the Higgs mass. As a rule of
thumb, the predictions for the top mass vary by about plus or minus 20 GeV
when $m_H$ is changed to 1 TeV or 60 GeV, respectively. This does
not apply to the $Z$ width into $b$ quarks and obviously the CDF result both of
which are insensitive to $m_H$. The quoted confidence levels would also be
different when groups of measurements such as $\nu N$ scattering were
broken down into the individual experiments. Also note that inclusion of the
uncertainty in the hadronic vacuum polarization would have little effect
on our conclusions.

\begin{table*}[hbt]
\setlength{\tabcolsep}{1.5pc}
\newlength{\digitwidth} \settowidth{\digitwidth}{\rm 0}
\catcode`?=\active \def?{\kern\digitwidth}
\caption{Various combined fits to the data sets described in the text.
The last column shows the standard deviations relative to the SLD result.
In the fit results and in the quoted errors the theoretical uncertainty
from the hadronic vacuum polarization (about 0.0003 in $\sin^2\th _{\rm eff}$)
is ignored.}
\label{tabfits}
\begin{tabular*}{\textwidth}{@{}l@{\extracolsep{\fill}}rlll}
\hline
                 & \multicolumn{1}{c}{$\sin^2\th _{\rm eff}^{\rm lept}$}
                 & \multicolumn{1}{c}{$m_t$ [GeV]}
                 & \multicolumn{1}{c}{$\chi^2$/ d.o.f.}
                 & \multicolumn{1}{c}{$\sigma$ (c.l.)} \\
\hline
LEP asymmetries & $ 0.2321\; (5)$ & $176 \pm 15$ & 11.2/7 &             \zw
plus '93 SLD    & $ 0.2315\; (4)$ & $193 \pm 12$ & 18.6/8 & 2.7 (99.4\%)\zw
plus all SLD    & $ 0.2316\; (4)$ & $192 \pm 12$ & 19.7/9 & 2.5 (98.6\%)\zw
\hline
all LEP         & $ 0.2323\; (4)$ & $169 \pm 12$     & 11.8/8 &             \zw
plus '93 SLD    & $ 0.2319\; (4)$ & $183^{+10}_{-11}$& 21.0/9 & 3.0 (99.8\%)\zw
plus all SLD    & $ 0.2319\; (4)$ & $182 \pm 11$     & 22.0/10& 2.8 (99.4\%)\zw
\hline
all but SLD \& CDF & $ 0.2324\; (3)$ & $166 \pm 11$  & 12.2/10&             \zw
plus '93 SLD    & $ 0.2321\; (3)$ & $177^{+10}_{-9}$ & 22.3/11& 3.2 (99.9\%)\zw
plus all SLD    & $ 0.2321\; (3)$ & $177^{+9}_{-10}$ & 23.3/12& 2.8 (99.6\%)\zw
\hline
all but SLD     & $ 0.2323\; (3)$ & $169 \pm 9$      & 12.4/11&             \zw
plus '93 SLD    & $ 0.2321\; (3)$ & $177^{+8}_{-9}$  & 22.4/12& 3.2 (99.8\%)\zw
plus all SLD    & $ 0.2321\; (3)$ & $176 \pm 8$      & 23.3/13& 2.8 (99.6\%)\zw
\hline
\end{tabular*}
\end{table*}

Independent of these subtleties is the fact that the fits deteriorate
substantially when the left-right asymmetry is included. There is generally
a discrepancy of slightly more or less than 3 $\sigma$ depending of whether
the 1992 data of the SLD collaboration are included.

\section{New physics?}

Table~\ref{tabfits} shows that in the case where only asymmetry measurements
are considered the discrepancy is
a little smaller, but still at least 2.5 $\sigma$. This is because the
asymmetries give the largest value for $m_t$ and the error is relatively large.
Of course it is here
that a direct comparison between LEP and SLC is possible and independent
of the Higgs mass and hadronic uncertainties. Thus, if one tries to explain
the discrepancy between LEP and SLC by new physics effects one should focus
on the asymmetries.

The processes at LEP 1 and SLC experiments are overwhelmingly dominated
by $Z$ exchange. Hence it is possible to use improved Born formulas whereby
new physics shows up as changes in the effective fermion couplings
to the $Z$. The kinds of new physics which might pop up that way include
\begin{itemize}
\item extended gauge structures (e.g. $Z^\prime$s),
\item non standard fermions (such as mirror fermions),
\item non standard Higgs structures,
\item heavy new particles in loops (like supersymmetry).
\end{itemize}

\begin{table*}[hbt]
\setlength{\tabcolsep}{1.5pc}
\catcode`?=\active \def?{\kern\digitwidth}
\caption{The effects of new physics sensitive to $A_{LR}$ on $M_W$ and other
observables.
See reference [12] for details about the types of new
physics shown here.}
\label{tabnp}
\begin{tabular*}{\textwidth}{@{}l@{\extracolsep{\fill}}cll}
\hline
                  \multicolumn{1}{c}{new physics}
                 & \multicolumn{1}{c}{increase of $M_W$ [MeV] when}
                 & \multicolumn{1}{c}{other effects (present error)} \\
 & increasing $A_{LR}$ by 0.02. &  \vier
\hline
\multicolumn{1}{c}{$Z^\prime$'s} \vier
\cline{1-1}
$Z_\Psi$ ($\sqrt{\frac{2}{3}}$)   &  420 &        \vier
$Z_\Psi$ ($-\sqrt{\frac{2}{3}}$)  &  470 & $\Gamma_{l^+ l^-}: + 2.2$ MeV
($\pm: 0.18$ MeV) \vier
$Z_\eta$ ($-\sqrt{\frac{1}{15}}$) &  135 & $A_{FB}^0(b): + 0.014$
($\pm: 0.0045$) \vier
$Z_\eta$ ($\sqrt{\frac{16}{15}}$) & 1070 & \vier
$Z_{3R}$ ($\sqrt{\frac{3}{5}}$)   &  380 & \vier
$Z_{3R}$ ($\sqrt{\frac{2}{3}}$)   &  730 & $\Gamma_{l^+ l^-}: + 6$ MeV \vier
\hline
nonstandard Higgs fields        &  445 & \vier
\hline
\multicolumn{1}{c}{extra fermions} \vier
\cline{1-1}
$e_R$-doublet                     &  --- & $A_{FB}^0(b): + 0.014$ \vier
$\nu_{e_L}$-singlet               &  135 & $\Gamma_Z: +20.6$ MeV
($\pm: 3.8$ MeV) \vier
$\mu_L$-singlet                   &  135 & $\Gamma_Z: +22.9$ MeV \vier
$\nu_{\mu_L}$-singlet             &  135 & (see $\nu_{e_L}$) \vier
\hline
\multicolumn{1}{c}{new physics in loops} \vier
\cline{1-1}
$T$ parameter                     &  445 & \vier
$S$ parameter                     &  205 & $A_{FB}^0(b): + 0.014$ \vier
$m_t$                             &  530 & \vier
\hline
\end{tabular*}
\end{table*}

As noted above, the common fit to the LEP asymmetries is relatively poor
and consistent only at the 13.1\% c.l. The largest $\chi^2$ contribution
comes from $A_{FB} (\tau)$, which shows a 2.7 $\sigma$ deviation w.r.t.
other LEP asymmetries. Its removal would render the fit perfect.
A fit to the leptonic forward-backward asymmetries alone yields a
\kaid of 5.2/2. The universality test at LEP is usually referred
to as excellent, but this fit excludes lepton universality at the 92.6\% c.l.
(Of course, the universality
test w.r.t. the axial vector couplings as determined from the widths is
indeed excellent.) Thus, one might try to give up lepton universality in order
to improve the fit.
\subsection{No SLC data}
A common fit to the LEP asymmetries without the SLD result and without
assuming lepton universality gives
\be
\ba{rcrcl}
   \sin^2\th _{\rm eff}^{\rm e} ({\rm LEP}) &=& 0.2325 &\pm& 0.0006 , \vier
   \sin^2\th _{\rm eff}^{\rm \mu} ({\rm LEP}) &=& 0.2327 &\pm& 0.0026 , \vier
   \sin^2\th _{\rm eff}^{\rm \tau} ({\rm LEP}) &=& 0.2301 &\pm& 0.0012 ,
\ea
\ee
with a \kaid of 7.8/5, which means only a modest improvement.
The goodness of fit (g.o.f.) increased from 13.1\% to 16.7\%.
Here, the largest \kai contribution (5.1) comes from a ``new discrepancy''
between the effective tau angles as determined from
\be
\ba{rcl}
   A_{FB} (\tau) &\Rightarrow&
    \sin^2\theta_{\rm eff}^{\tau} = 0.2228 \pm 0.0035
\ea
\ee
and
\be
\ba{rcl}
  A_\tau ({\cal P}_\tau)  &\Rightarrow&
  \sin^2\theta_{\rm eff}^{\tau} = 0.2311 \pm 0.0013 .
\ea
\ee
This amounts to 2.3 $\sigma$!

\subsection{Including 1993 SLC data}
Including the left-right asymmetry we find
\be
\ba{rcrcl}
   \sin^2\th _{\rm eff}^{\rm e} ({\rm LEP}) &=& 0.2316 &\pm& 0.0005 , \vier
   \sin^2\th _{\rm eff}^{\rm \mu} ({\rm LEP}) &=& 0.2335 &\pm& 0.0025 , \vier
   \sin^2\th _{\rm eff}^{\rm \tau} ({\rm LEP}) &=& 0.2302 &\pm& 0.0012 ,
\ea
\ee
where the g.o.f.\ actually deteriorated from 1.7\% to 1.1\%. The discrepancy
between SLC and LEP increased from 2.7 to 3.0 $\sigma$. Giving up lepton
universality would thus worsen the situation.

\subsection{Arbitrary fermion couplings}
If one wishes to allow for more radical types of new physics, one can also
leave all fermion couplings as free parameters and see whether a decent fit
can be obtained. For that matter we have to leave out the
charge asymmetry, since in their analysis LEP physicists had to
assume the validity of the SM (for the quark sector!).

The \kai value of 13.3/3 virtually excludes this possibility at the 99.6\% c.l.
Moreover, radical choices of the mixing angles have to be made,
\be
\ba{rcl}
  \sin^2\theta_{\rm eff}^e      &=& 0.2307 \pm 0.0007, \\
  \sin^2\theta_{\rm eff}^{\mu}  &=& 0.2343 \pm 0.0024, \\
  \sin^2\theta_{\rm eff}^{\tau} &=& 0.2303 \pm 0.0012, \\
  \sin^2\theta_{\rm eff}^c      &=& 0.2435 \pm 0.0256, \\
  \sin^2\theta_{\rm eff}^b      &=& 0.3402 \pm 0.0452.
\ea
\ee
In fact, the above value of $\sin^2\theta_{\rm eff}^b$ corresponds to
$A_b = 0.84 \pm 0.05$, which does not even agree with the direct measurement
of $A_b = 0.99 \pm 0.13$ from SLD.

\subsection{A closer look at new physics}
We have seen that a statistical fluctuation is as unlikely to explain the
observed discrepancy
as is new physics. If the SLD measurement is accurate some kind of
new physics must explain why the left-right asymmetry is so high.
At the same time there must be a conflict with other
experimental results. In table~\ref{tabnp} I list some kinds of new physics
which are sensitive to $A_{LR}$. The results shown were obtained with help
of reference~\cite{LLM}.

It turns out that virtually all kinds of new physics which are sensitive to
$A_{LR}$ simultaneously induce a shift in the $W$ mass. This correlation with
$M_W$ does not occur for an extra right handed lepton doublet which mixes
with the electron. Still, this possibility would require that
several of the LEP asymmetry measurements are wrong,
most notably ${\cal P}_\tau^{FB}$ and $A_{FB}(b)$.

\section{Conclusions}
The new determination of the left-right asymmetry at SLC and the combined
asymmetry measurements at LEP are inconsistent with each other at the 99.4\%
confidence level.
Including previous SLC data gives an overall discrepancy of 2.5 $\sigma$.
A fit exclusively to LEP asymmetries is relatively poor.

On the other hand, if the Standard Model is assumed to be correct satisfactory
fits can be obtained when the SLD result is left out. The discrepancy of the
left-right asymmetry to such a global fit result is about 3 $\sigma$.

Although the test of universality w.r.t. the leptonic effective mixing
angles at LEP is relatively poor, an inclusion of the SLC result worsens
the fit. Even a parametrization of most general kinds of new physics does
not yield satisfying fit results. New physics can only explain the high
value of $A_{LR}$ under the simultaneous assumption that several other
experiments are wrong.

\section*{Acknowledgements}
This work was done in collaboration with Paul Langacker.
We are indebted to Bernd Kniehl who made his program for the calculation
of $\sin^2\th_{\overline {MS}}$ and $M_W$ available to us.

\end{document}